\documentclass[fp,twocolumn]{jpsj3}
\usepackage{graphicx}
\usepackage{amsmath,amssymb}
\usepackage{ascmac}
\usepackage{url}
\usepackage{layout}
\usepackage{color}
\usepackage{braket}
\usepackage{physics}
\usepackage{mathrsfs}
\usepackage{booktabs}
\usepackage{here}
\usepackage{listings}
\usepackage{dcolumn}
\usepackage{bm}

\def\lesssim{\ \raise.3ex\hbox{$<$}\kern-0.8em\lower.7ex\hbox{$\sim$}\ }
\def\gesim{\ \raise.3ex\hbox{$>$}\kern-0.8em\lower.7ex\hbox{$\sim$}\ }

\begin{document}
\title{Nuclear Spin-Lattice Relaxation Rate in Odd-Frequency Superconductivity}
\author{Shumpei Iwasaki$^*$ and Yoji Ohashi}
\inst{Department of Physics, Keio University, 3-14-1 Hiyoshi Kohoku-ku, Yokohama 223-8522, Japan}
\abst{We theoretically investigate the temperature dependence of nuclear spin-lattice relaxation rate $T_1^{-1}$ in bulk odd-frequency superconductivity. For a model odd-frequency pairing interaction, we first evaluate the superconducting order parameter, within the framework of the combined path-integral formalism with the saddle-point approximation. We then calculate $T_1^{-1}$ below the superconducting phase transition temperature $T_{\rm c}$, to see how the odd-frequency pairing affects this physical quantity. In the odd-frequency $p$-wave state, while the so-called coherence peak is suppressed as in the even-frequency $p$-wave case, $T_1^{-1}$ is found to exhibit the Korringa-law-like behavior ($T_1^{-1}\propto T$) except just below $T_{\rm c}$, even without impurity scatterings. In the odd-frequency $s$-wave case, the behavior of $T_{1}^{-1}$ is found to be sensitive to the detailed spin structure of the superconducting order parameter: In a case, $T_1^{-1}$ is enhanced far below $T_{\rm c}$, being in contrast to the conventional (even-frequency) $s$-wave BCS case, where the coherence peak appears just below $T_{\rm c}$. We also show that the calculated $T_1^{-1}$ in the odd-frequency $p$-wave case well explains the recent experiment on CeRh$_{0.5}$Ir$_{0.5}$In$_5$, where the possibility of odd-frequency $p$-wave superconductivity was recently suggested experimentally.}
\maketitle
\par
\section{Introduction}
\par
In condensed matter physics, odd-frequency superconductivity~\cite{Linder2019,Berezinskii1974} has extensively been discussed in various strongly correlated electron systems, such as heavy fermion compounds~\cite{Emery1992,Tsvelik1993,Tsvelik1994,Fuseya2003,Shigeta2012,Hoshino2014A,Hoshino2014B,Hoshino2016,Funaki2014,Otsuki2015}, organic conductors~\cite{Shigeta2009,Shigeta2011,Pratt2013,Fukui2018,Fukui2019,Yoshida2021}, as well as transition metal oxides~\cite{Balatsky1992,Abrahams1993,Matsubara2021,Inokuma2024}. Recently, this unconventional pairing state has also been discussed in cold Fermi gas physics\cite{Kalas2008,Arzamasovs2018,Linder2019,Iwasaki2024}. At present, the realization of bulk odd-frequency superconductivity is still under debate; however, various properties of this state, such as the superconducting transition temperature $T_{\rm c}$, the superconducting order parameter, as well as the specific heat, have already been examined theoretically~\cite{Kusunose2011A,Matsumoto2012,Kusunose2012}.
\par
Recently, Kawasaki and co-workers\cite{Kawasaki2020} experimentally reported an anomaly of the nuclear spin-lattice relaxation rate $T_1^{-1}$\cite{Parks,Hebel1960,Moriya1956,Moriya1963} near a quantum critical point (QCP) in heavy-fermion superconductor CeRh$_{0.5}$Ir$_{0.5}$In$_{5}$. Based on this, they pointed out that an odd-frequency $p$-wave superconducting state may be realized there\cite{Kawasaki2020}: In this material, antiferromagnetism and superconductivity coexist. When the magnetic phase is suppressed by applying the pressure, QCP appears, around which the spin-singlet pairing is believed to be favorable, due to a pairing interaction mediated by antiferromagnetic spin fluctuations\cite{Pfleiderer2008}. Indeed, under the pressure away from QCP to some extent, the power-low behavior of the spin-lattice relaxation rate ($T_1^{-1}\propto T^3$) is observed, implying the realization of even-frequency spin-singlet $d$-wave superconductivity~\cite{Fuseya2003,Sigrist1991,Bang2004}. On the other hand, the observed $T_1^{-1}$ very near QCP slightly decreases only just below $T_{\rm c}$, but  exhibits the linear-temperature dependence ($T_1^{-1}\propto T$) at low temperatures. Since this Korringa-law-like behavior~\cite{Korringa1950} implies the non-zero superconducting density of states (SDOS) around $\omega=0$, Ref.\cite{Kawasaki2020} suggested the realization of an odd-frequency spin-singlet $p$-wave pairing state, as a possible explanation for the observed anomaly\cite{note2}. 
\par
At present, however, it is unclear whether or not the odd-frequency $p$-wave state can really explain the observed Korringa-law-like behavior $T_1^{-1}$ in heavy-fermion superconductor CeRh$_{0.5}$Ir$_{0.5}$In$_{5}$\cite{Fuseya2003,Kawasaki2003,Zheng2004}. Thus, in the current stage of research, it is a crucial issue to theoretically assess the recent experimental suggestion\cite{Kawasaki2020}, by explicitly evaluating $T_1^{-1}$ in this unconventional pairing state. Since the gapless SDOS (as well as the linear temperature dependence of $T_1^{-1}$) can also be obtained in even-frequency non-$s$-wave superconductors with non-magnetic impurities\cite{Bang2004,Schmitt-Rink1986,Ishida1993,Ishida2000,Hotta2000,Puchkaryov1998,Ohashi2000,Yoshioka2012}, such a theoretical assessment would be useful to judge which scenario is more promising.
\par
Motivated by this circumstance, in this paper, we theoretically investigate $T_1^{-1}$ in bulk odd-frequency superconductivity\cite{note1,Tanaka2016,Tanaka2018}. Regarding this, we note that the following recent theoretical developments are important for our study: (1) When one simply extends the mean-field BCS Hamiltonian formalism to the bulk odd-frequency pairing state, the resulting odd-frequency superconducting state is known to become stable {\it above} $T_{\rm c}$\cite{Heid1995}. (Note that the superconducting state in CeRh$_{0.5}$Ir$_{0.5}$In$_{5}$ is, of course, observed below $T_{\rm c}$.) This difficulty was recently overcome by employing the combined path-integral approach with an appropriate saddle point approximation\cite{Solenov2009,Kusunose2011B,Fominov2015,note}. (2) One needs to carefully choose the frequency dependence of a model odd-frequency pairing interaction. Otherwise, SDOS is known to unphysically become {\it negative} around $\omega=0$\cite{Iwasaki2024}. This problem is particularly serious in calculating $T_1^{-1}$, because it is sensitive to SDOS around $\omega=0$. For this difficulty, the authors have recently found a model odd-frequency pairing interaction that guarantees the positivity of SDOS\cite{Iwasaki2024}. 
\par
Using these theoretical developments, we explicitly calculate $T_1^{-1}$ in the odd-frequency $p$-wave superconducting state, to examine to what extent the observed temperature dependence of this quantity in CeRh$_{0.5}$Ir$_{0.5}$In$_{5}$ can be explained by this unconventional pairing state. Besides this, we also deal with the odd-frequency $s$-wave case, to see how the so-called coherence peak appearing just below $T_{\rm c}$ in the conventional (even-frequency) $s$-wave BCS state is affected by the odd-frequency pairing. 
\par
This paper is organized as follows: In Sec. 2, we explain our formulation to evaluate $T_1^{-1}$ in the odd-frequency superconducting state. We show our results in Sec. 3. Throughout this paper, we set $\hbar=k_{\rm B}=1$, and the system volume $V$ is taken to be unity, for simplicity.
\par
\par
\section{Formulation}
\par
\subsection{Model odd-frequency superconductivity}
\par
Following Refs.~\cite{Solenov2009,Kusunose2011B}, we start from the partition function in the path-integral representation:
\begin{equation}
Z=\int {\cal D}\bar{\psi }{\cal D}\psi e^{-S[\bar{\psi},\psi]}
=\prod _{k,\sigma}\int d\bar{\psi }_{k,\sigma} d\psi _{k,\sigma}
e^{-S[\bar{\psi},\psi]},
\label{eq.1}
\end{equation}
where the Grassmann variable $\psi _{k,\sigma}$ and its conjugate $\bar{\psi }_{k,\sigma}$ describe the electron field with spin $\sigma=\uparrow,\downarrow$. In Eq.~(\ref{eq.1}), we have introduced the abbreviated notation $k=({\bm k},i\omega_n)$, where $\omega_n$ is the fermion Matsubara frequency. The action $S[\bar{\psi},\psi]=S_0+S_1$ consists of the kinetic term $S_0$ and the pairing interaction term $S_1$, that are given by, respectively,
\begin{equation}
S_0 =\sum _{k,\sigma}\bar{\psi }_{k,\sigma}
\left[-i\omega_n +\xi _{\bm k}\right] \psi _{k,\sigma},
\label{eq.2}
\end{equation}
\begin{equation}
S_{1} =\frac{1}{\beta }\sum _{q,k,k'}{\textstyle V_l\left(k+\frac{q}{2},k'+\frac{q}{2}\right)}\overline{\psi }_{k+q,\uparrow}\overline{\psi }_{-k,\downarrow} \psi _{-k',\downarrow} \psi _{k'+q,\uparrow}.
\label{eq.3}
\end{equation}
Here, $\beta=1/T$ is the inverse temperature, and $q=({\bm q},i\nu_m)$ with $\nu_n$ being the boson Matsubara frequency. In Eq.~(\ref{eq.2}), $\xi_{\bm k}=\varepsilon_{\bm k}-\varepsilon_{\rm F}={\bm k}^2/(2m)-\varepsilon_{\rm F}$ is the kinetic energy of an electron, measured from the Fermi energy $\varepsilon_{\rm F}=k_{\rm F}^2/(2m)$ (where $k_{\rm F}$ is the Fermi momentum and $m$ is an electron mass). $V_l(k,k')=V_l({\bm k},i\omega_n,{\bm k}',i\omega_n')$ in Eq.~(\ref{eq.3}) is a pairing interaction, where the $\omega_n$- and $\omega_n'$-dependence describe retardation effects of this interaction. The subscript `$l$' in $V_l(k,k')$ specifies the pairing types that we are considering in this paper: (1) odd-frequency spin-singlet $p$-wave pairing ($l={\rm odd},p$), (2) odd-frequency spin-triplet ($S_z=0$) $s$-wave pairing ($l={\rm odd},s$), and (3) conventional even-frequency spin-singlet $s$-wave BCS state ($l={\rm even},s$). 
\par 
Regarding the type (2), we note that we deal with the spin triplet with $\ket{S,S_z}=\ket{1,0}$ in this section to simply explain our formulation (which we call ``$ \uparrow\downarrow + \downarrow\uparrow $ state'' in what follows). In Sec. 3, however, we also consider the case with the mixture of $\ket{S,S_z}=\ket{1,\pm 1}$ for comparison, which is reffered to as ``$ \uparrow\uparrow + \downarrow\downarrow $ state'' in the following discussions. Since the formulation for the latter is essentially the same as the former, we summarize its outline in Appendix B.
\par
In this paper, we do not discuss the origin of the pairing interaction $V_l(k,k')$ in Eq. (\ref{eq.3}), but simply assume the following separable form~\cite{Iwasaki2024}:
\begin{equation}
V_l(k,k') =-U_l\gamma_l({\bm k}, i\omega_n) \gamma_l({\bm k}', i\omega_n'),
\label{eq.4}
\end{equation}
where $-U_l~(<0)$ is a coupling strength. An advantage of this model interaction is that the ${\bm k}$- and $\omega_n$-dependence of the superconducting order parameter $\Delta_l({\bm k},i\omega_n)$ are immediately determined by the basis function $\gamma_l({\bm k}, i\omega_n)$ as,
\begin{equation}
\Delta_{l}({\bm k},i\omega_n)=\Delta_l\gamma_l({\bm k},i\omega_n),
\label{eq.5}
\end{equation}
where $\Delta_l$ is a constant. Keeping this in mind, we choose the basis functions for the above-mentioned three pairing types as
\begin{align}
\gamma_{{\rm odd},p}({\bm k},i\omega_n)
&=\cos(\theta_{\bm k})
\frac{\omega_n}
{|\omega_n|}
\frac{\sqrt{\omega_n^2 +\xi_{\bm k}^2}}
{\sqrt{\omega_n^2+\xi_{\bm k}^2+\Lambda^2}},
\label{eq.7}
\\
\gamma_{{\rm odd},s}({\bm k},i\omega_n)
&=
\frac{\omega_n}
{|\omega_n|}
\frac{\sqrt{\omega_n^2 +\xi_{\bm k}^2}}
{\sqrt{\omega_n^2+\xi_{\bm k}^2+\Lambda^2}},
\label{eq.6}
\\
\gamma_{{\rm even},s}({\bm k},i\omega_n)
&=
1.
\label{eq.8}
\end{align}
Among these, Eqs. (\ref{eq.7}) and (\ref{eq.6}) are odd functions with respect to $\omega_n$, where $\Lambda$ physically describes retardation effects of the pairing interaction in Eq. (\ref{eq.4})\cite{Iwasaki2024}. In Eq. (\ref{eq.7}) (odd-frequency $p$-wave case), we have chosen the polar state being characterized by $\cos(\theta_{\bm k})$, as an example of $p$-wave symmetry~\cite{Zheng2004}. 
\par
We note that one might consider simpler odd-frequency basis functions than Eqs.  (\ref{eq.7}) and (\ref{eq.6}), e.g., $\gamma_{{\rm odd},p}({\bm k},i\omega_n)=\cos(\theta_{\bm k})\omega_n$, $\gamma_{{\rm odd},s}({\bm k},i\omega_n)=\omega_n$, and Eqs. (\ref{eq.7}) and (\ref{eq.6}) with $\xi_{\bm k}=0$. However, as pointed out in Ref. \cite{Iwasaki2024}, these unphysically give negative SDOS around $\omega=0$. On the other hand, we will see in Sec. 3 that the basis functions in Eqs.~(\ref{eq.7}) and (\ref{eq.6}) safely give positive SDOS. Because SDOS around $\omega=0$ is crucial for $T_1^{-1}$, choosing an appropriate model interaction that satisfies this required condition is important for our purpose.
\par
We introduce the Cooper-pair Bose field $\Phi$ (as well as its conjugate field ${\bar \Phi}$) by the Stratonovich-Hubbard transformation~\cite{Stratonovich1958,Hubbard1959}. The resulting partition function has the form,
\begin{equation}
Z\propto
\int{\cal D}{\bar \Phi}{\cal D}\Phi
\int{\cal D}{\bar \psi}{\cal D}\psi 
e^{-S[{\bar \psi},\psi,{\bar \Phi},\Phi]},
\label{eq.9}
\end{equation}
where
\begin{align}
&~~~~~~~~\int{\cal D}{\bar \Phi}{\cal D}\Phi 
=\prod_q \int d{\rm Re}[\Phi_q] d{\rm Im}[\Phi_q],
\label{eq.10}
\\
&S[{\bar \psi},\psi,{\bar \Phi},\Phi]=
S_0+
\sum_q
\left[
{{\bar \Phi}_q \Phi_q \over U}
-{\bar \rho}_q \Phi_q
-\rho _q{\bar \Phi}_q
\right],
\label{eq.11}
\\
&\rho_q=
\frac{1}{\sqrt{\beta}}
\sum_k 
\gamma_l\left({\bm k}+\frac{{\bm q}}{2},i\omega_n + \frac{i\nu_m}{2} \right)
\psi_{-k,\downarrow}
\psi_{k+q,\uparrow},
\label{eq.12}
\\
&{\bar \rho}_q=
\frac{1}{\sqrt{\beta}}
\sum_k
\frac{1}{\sqrt{\beta}}
\gamma_l\left({\bm k}+\frac{{\bm q}}{2},i\omega_n + \frac{i\nu_m}{2} \right)
{\bar \psi}_{k+q,\uparrow}
{\bar \psi}_{-k,\downarrow}.
\label{eq.13}
\end{align}
In this paper, we treat the superconducting state within the mean-field level. In the present path-integral scheme, it corresponds to employing the saddle-point approximation, where the path-integrals with respect to ${\bar \Phi}$ and $\Phi$ in Eq.~(\ref{eq.9}) are replaced by the values at the saddle-point solution~\cite{Tempere2012}. Choosing the Cooper-pair fields at the saddle point as~\cite{Solenov2009,Kusunose2011B},
\begin{align}
\left\{
\begin{array}{l}
\Phi_q =\sqrt{\beta } \Delta_l \delta_{q,0},\\
{\bar \Phi}_q =\sqrt{\beta }\Delta^*_l \delta_{q,0},
\end{array}
\right.
\label{eq.22}
\end{align}
and introducing the Nambu spinor, $\Psi_k =( \psi _{k,\uparrow} ,\psi _{-k,\downarrow})^{\mathrm{T}}$, $\Psi_k^\dagger =( \bar{\psi}_{k,\uparrow} ,\psi _{-k,\downarrow})$, we obtain the saddle-point partition function~\cite{Randeria2008,Stoof2009}
\begin{equation}
Z_{\mathrm{SP}} =\int \mathcal{D}\overline{\psi }\mathcal{D} \psi e^{-S_{\mathrm{SP}} [\overline{\psi } ,\psi ]} .
\label{eq.b1}
\end{equation}
Here,
\begin{equation}
S_{\mathrm{SP}} =\sum _{k} \Psi _{k}^{\dagger }\left[-\hat{G}_{l}^{-1}( k)\right] \Psi _{k} +\beta \sum _{\bm k} \xi _{\bm k} +\frac{\beta |\Delta _{l} |^{2}}{U},
\label{eq.b2}
\end{equation}
is the saddle-point action $S_{\rm SP}$, where
\begin{align}
{\hat G}_{l}(k)
=&
{1 \over 
i\omega_n-\xi_{\bm k}\tau_3+
\begin{pmatrix}
0 & 
\Delta_l \gamma_l({\bm k},i\omega_n) \\
\Delta_l^* \gamma_l({\bm k},i\omega_n) &
0
\end{pmatrix}
}
\nonumber
\\
=&
-{1 \over \omega_n^2+\xi_{\bm k}^2+|\Delta_l\gamma_l({\bm k},i\omega_n)|^2}
\nonumber
\\
&\times
\begin{pmatrix}
i\omega_n+\xi_{\bm k} & 
-\Delta_l \gamma_l({\bm k},i\omega_n) \\
-\Delta_l^* \gamma_l({\bm k},i\omega_n) &
i\omega_n-\xi_{\bm k}
\end{pmatrix},
\label{eq.21}
\end{align}
is the $2\times 2$ matrix mean-field single-particle thermal Green's function~\cite{Iskin2005,Iskin2006A,Iskin2006B,Iskin2006C,Cao2013,Tempere2008}, with $\tau_{i=1,2,3}$ being the Pauli matrices acting on particle-hole space. Executing the fermion path-integrals in Eq.~(\ref{eq.b1})~\cite{Tempere2012}, one obtains the following mean-field thermodynamic potential~\cite{Iwasaki2024}:
\begin{align}
\Omega_{\rm MF}
&=
-T\ln Z_{\rm SP}
\nonumber
\\
=&
{|\Delta_l|^2 \over U}
+\sum_{\bm k}\xi_{\bm k}
\nonumber
\\
&-{1 \over \beta}\sum_k
\ln\left[\omega_n^2+\xi_{\bm k}^2+|\Delta_l|^2\gamma_l({\bm k},i\omega_n)^2\right].
\label{eq.24}
\end{align} 
The mean-field superconducting order parameter $\Delta_l$ in Eq.~(\ref{eq.22}) is determined from the saddle point condition,
\begin{equation}
{\partial \Omega_{\rm MF} \over \partial \Delta_l^*}=0.
\label{eq.23}
\end{equation}
Substituting Eq.~(\ref{eq.24}) into Eq.~(\ref{eq.23}), we obtain the BCS-type gap equation,
\begin{align}
1
&=
{U \over \beta}
\sum_{{\bm k},\omega_n}
{\gamma_l({\bm k},i\omega_n)^2 \over
\omega_n^2+\xi_{\bm k}^2+|\Delta_l|^2\gamma_l({\bm k},i\omega_n)^2}
\nonumber
\\
&=U\sum _{\bm k}
{\eta_l({\bm k})^2 \over 2E_{l}({\bm k})}
\tanh\left({E_{l}({\bm k}) \over 2T}\right),
\label{eq.25}
\end{align}
where $(\eta_{{\rm odd},p}({\bm k}),\eta_{{\rm odd},s}({\bm k}),\eta_{{\rm even},s}({\bm k}))=(\cos(\theta_{\bm k}),1,1)$, and 
\begin{align}
E_{{\rm odd},p}({\bm k})&=\sqrt{\xi_{\bm k}^2+\Lambda^2+|\Delta_{{\rm odd},p}|^{2}\cos^2(\theta_{\bm k})},
\label{eq.26} 
\\
E_{{\rm odd},s}({\bm k})&=\sqrt{\xi_{\bm k}^2+\Lambda^2+|\Delta_{{\rm odd},s}|^{2}},
\\
E_{{\rm even},s}({\bm k})&=\sqrt{\xi_{\bm k}^2+|\Delta_{{\rm even},s}|^{2}},
\label{eq.26b}
\end{align}
describe Bogoliubov single-particle excitations. We briefly note that the gap equation~(\ref{eq.25}) can also be obtained from the off-diagonal component $G_l^{(1,2)}$ of the $2\times 2$ matrix Green's function in Eq.~(\ref{eq.21}) as,
\begin{equation}
\Delta_l\gamma({\bm k},i\omega_n)=
{1 \over \beta}\sum_{k'}V_{k,k'}G_l^{(1,2)}(k').
\label{eq.27}
\end{equation}
\par
As pointed out in Refs.~\cite{Solenov2009,Kusunose2011B}, the choice in Eq.~(\ref{eq.22}) guarantees the expected thermodynamic behavior that the superconducting state is stable {\it below} $T_{\rm c}$. Indeed, the gap equation (\ref{eq.25}) gives $\Delta_l\ne 0$ only below $T_{\rm c}$ which is determined from Eq. (\ref{eq.25}) with $\Delta_l=0$\cite{note4}. In contrast, the Hamiltonian formalism leads to the different case that the second equation in Eq.~(\ref{eq.22}) is replaced by ${\bar \Phi}_q=-\sqrt{\beta}\Delta_l^*\delta_{q,0}$~\cite{Solenov2009,Kusunose2011B}, leading to the unphysical situation, as mentioned previously.
\par
In computations, we actually replace the ${\bm k}$-summation in the gap equation (\ref{eq.25}) with the $\xi_{\bm k}$- and $\theta_{\bm k}$-integrations as
\begin{equation}
\sum_{\bm k}\to
{1 \over 2}\rho_{\rm N}(0)
\int_0^\pi d\theta_{\bm k}\sin\theta_{\bm k}
\int_{-\omega_{\rm c}}^{\omega_{\rm c}}d\xi_{\bm k},
\label{eq.comp}
\end{equation}
where $\omega_{\rm c}$ is a cutoff energy describing the energy region where the pairing interaction works ($T_{\rm c}\ll\omega_{\rm c}\ll\varepsilon_{\rm F}$). In Eq. (\ref{eq.comp}), we have approximated the normal-state density of state (NDOS) $\rho_{\rm N}(\omega)$ to the value $\rho_{\rm N}(0)=mk_{\rm F}/(2\pi^2)$ at the Fermi level, by assuming that the region near the Fermi surface is important.
\par
\begin{figure}[t]
\centering
\includegraphics[width=0.4\textwidth]{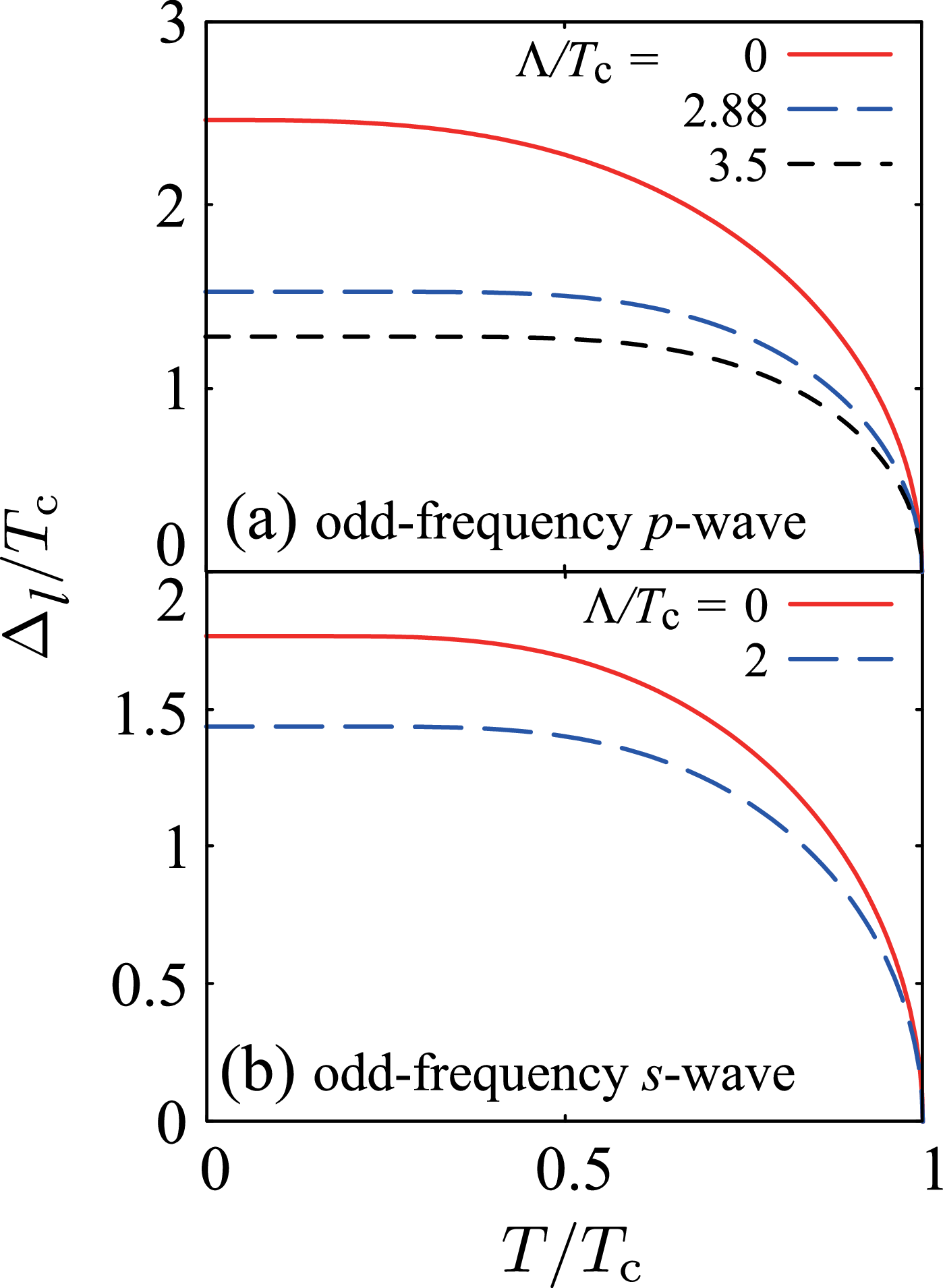}
\caption{Calculated superconducting order parameter $\Delta_l$ as a function of the temperature. (a) Odd-frequency $p$-wave state. (b) Odd frequency $s$-wave state.}
\label{fig1}
\end{figure}

We show the calculated superconducting order parameter $\Delta_l$ in Fig. \ref{fig1}\cite{note4}, which will be used in computing $T_1^{-1}$ in Sec.III. We briefly note that, when $\Lambda=0$ (where $\Delta_l({\bm k},i\omega_n)=\Delta_l\eta_l({\bm k})\times{\rm sign}(\omega_n)$, the temperature dependence of the order parameter in the odd-frequency $p$-wave state (odd-frequency $s$-wave state) is the same as that in the ordinary even-frequency $p$-wave polar state (even-frequency $s$-wave BCS state), because the gap equation (\ref{eq.25}) only depends on $|\Delta_l({\bm k},i\omega_n)|^2$. 
\par
\par
\par
\subsection{Spin-lattice relaxation rate $T_1^{-1}$}
\par
The spin-lattice relaxation rate $T_1^{-1}$ is given by\cite{Moriya1956,Moriya1963,Tanaka2018,Nagai2008}
\begin{equation}
\left[ TT_{1,l}( T)\right]^{-1} =C\lim _{\nu \rightarrow 0}\sum _{\bm q}
\frac{\mathrm{Im} \chi^{-+}_{l}\left({\bm q} ,i\nu_{m}\rightarrow \nu +i\delta \right)}{\nu },
\label{eq.a1}
\end{equation}
where $C$ is a constant, being related to the hyperfine coupling between nuclear-spin and electron-spin, and $\delta$ is an infinitesimally small positive number. In Eq. (\ref{eq.a1}),
\begin{align}
\chi^{-+}_{l}({\bm q} ,i\nu _{m}) & =\langle S_{-}(q) S_{+}(-q) \rangle 
\nonumber
\\
&=
\sum _{k,k'} \langle \overline{\psi }_{k,\downarrow} \psi _{k+q,\uparrow}\overline{\psi }_{k'+q,\uparrow} \psi _{k',\downarrow} \rangle,
\label{eq.a2}
\end{align}
is the transverse dynamical spin susceptibility in the Matsubara formalism, where 
\begin{align}
S_{-}(q)=\sum _{k}\overline{\psi }_{k,\downarrow} \psi _{k+q,\uparrow},
\\
S_{+}(-q)=\sum _{k}\overline{\psi }_{k+q,\uparrow} \psi _{k,\downarrow}.
\end{align}
\par
Evaluating Eq. (\ref{eq.a1}) within the mean-field theory explained in Sec. 2.2, we obtain
\begin{align}
&
\frac
{\left[ TT_{1,l}( T)\right]^{-1}}
{\left[ T_{\rm c}T_{1,l}(T_{\rm c})\right]^{-1}}
\nonumber
\\
&= 
\frac{\beta}{4\rho_0^2}
\int _{-\infty }^{\infty } d\omega 
\left[ \rho _{{\rm S},l}^2(\omega) 
+
|\Delta_l|^2\rho_{\Delta,l}(\omega )^2
\right]
{\rm sech}^2
\left({\beta\omega \over 2}\right).
\label{eq.x1}
\end{align}
We summarize the derivation in Appendix A. In Eq. (\ref{eq.x1}), 
\begin{equation}
\rho_{{\rm S},l}(\omega)=\left. -\frac{1}{\pi}
\sum_{\bm k}{\rm Im}
\left[
G_{l}^{( 1,1)}\left({\bm k},i\omega _{n}\rightarrow \omega +i\delta \right)
\right]\right| _{\omega _{n}>0},
\label{eq.x2}
\end{equation}
is just SDOS, so that we call the first term in $[\cdot\cdot\cdot]$ in Eq. (\ref{eq.x1}) the `SDOS term' in what follows. Since the second term in $[\cdot\cdot\cdot]$ in Eq. (\ref{eq.x1}) is related to the coherence effect in the conventional (even-frequency) $s$-wave BCS state, we call it the coherence term. When the pairing symmetry in momentum space is the $s$-wave type, $\rho_{\Delta,l}(\omega)$ involved in the coherence term is given by
\begin{equation}
\rho_{\Delta,{\rm odd},s}(\omega)=\left. 
\frac{1}{\pi}
\sum_{\bm k}{\rm Re}
\left[F_{{\rm odd},s}
({\bm k},i\omega_n\rightarrow \omega +i\delta)
\right]\right|_{\omega_n>0},
\label{eq.x4}
\end{equation}
\begin{equation}
\rho_{\Delta,{\rm even},s}(\omega)=\left. 
\frac{1}{\pi}
\sum_{\bm k}
{\rm Im} 
\left[
F_{{\rm even},s}
({\bm k},i\omega_n\rightarrow \omega +i\delta)
\right]\right|_{\omega_n>0},
\label{eq.x3}
\end{equation}
where $F_l({\bm k},i\omega_m)$ is related to the off-diagonal component of the Nambu Green's function in Eq. (\ref{eq.21}) as
\begin{equation}
G_l^{(1,2)}({\bm k},i\omega_m)=\Delta_lF_l({\bm k},i\omega_m).
\label{eq.xx4}
\end{equation}
In the odd-frequency $p$-wave case, on the other hand, the coherence term in Eq. (\ref{eq.x1}) vanishes, because
\begin{align}
\rho_{\Delta,{\rm odd},p}(\omega)
&=\left. 
\frac{1}{\pi}
\sum_{\bm k}{\rm Re}
\left[F_{{\rm odd},p}
({\bm k},i\omega _{n}\rightarrow \omega +i\delta)
\right]
\right| _{\omega_n>0}
\nonumber
\\
&=0,
\label{eq.x5}
\end{align}
due to the sign change of the $p$-wave symmetry ($\Delta_{{\rm odd},p}({\bm k},i\omega_n)\propto\cos(\theta_{\bm k})$) in momentum space. We briefly note that this vanishing coherence term also occurs in the case of ordinary even-frequency spin-triplet $p$-wave superconductivity.
\par
In numerically calculating Eqs. (\ref{eq.x2})-(\ref{eq.x3}), we also employ the prescription in Eq. (\ref{eq.comp}) where we set $\omega_{\rm c}\to\infty$ (because the $\xi_{\bm k}$-integration well converges in these equations).
\par
\begin{figure}[t]
\centering
\includegraphics[width=0.4\textwidth]{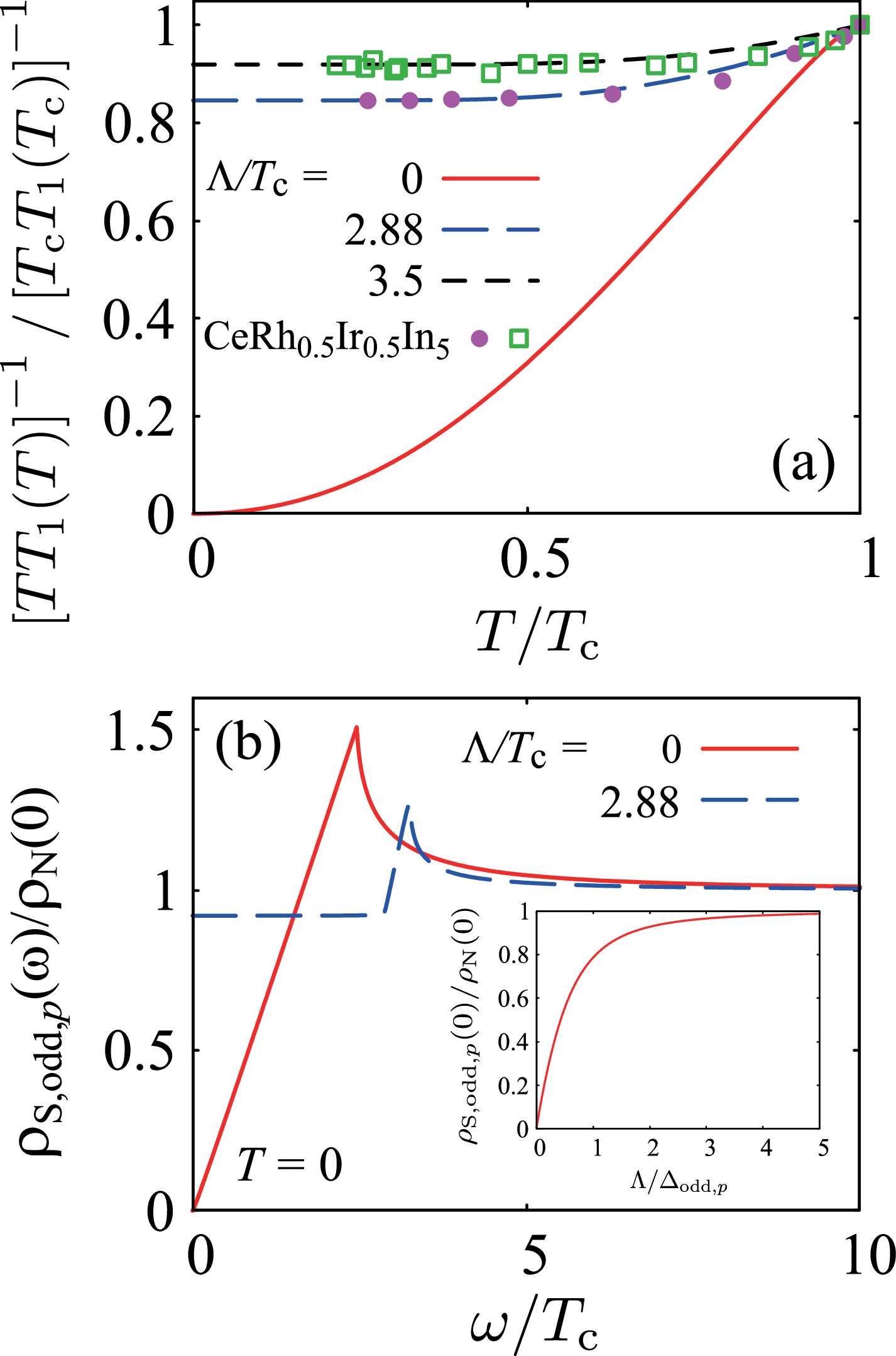}
\caption{(a) Calculated $(T_1T)^{-1}$ as a function of the temperature, in the odd-frequency $p$-wave superconducting state. The circle and square are the experimental data measured in CeRh$_{0.5}$Ir$_{0.5}$In$_5$ at 1.12GPa and 1.32 GPa, respectively~\cite{Kawasaki2020}. We set $\delta/T_{{\rm c}} = 10^{-2}$, which is also used in Figs. \ref{fig3} and \ref{fig4}. (b) SDOS $\rho_{{\rm S},{\rm odd},p}(\omega)$ at $T=0$. The inset shows $\rho_{{\rm S},{\rm odd},p}(0)$ as a function of $\Lambda/|\Delta_{{\rm odd},p}|$ at $T=0$.}
\label{fig2}
\end{figure}
\par
\section{Temperature dependence of spin-lattice relaxation rate $T_1^{-1}$ in odd-frequency superconductivity}
\par
\subsection{Odd-frequency $p$-wave state}
\par
Figure \ref{fig2}(a) shows $(T_1T)^{-1}$ as a function of the temperature in the odd-frequency $p$-wave superconducting state. In this state, since the coherence term in Eq. ({\ref{eq.x1}) vanishes, $T_1^{-1}$ is dominated by the SDOS term, where $\rho_{{\rm S},{\rm odd},p}(\omega)$ in this state has the form,
\begin{align}
\rho_{{\rm S},{\rm odd},p}(\omega)
=&
\sum_{\bm k}
{\Lambda^2 \over \Lambda^2+|\Delta_{{\rm odd},p}|^2 \cos^2(\theta_{\bm k})}
\delta(\omega-\xi _{\bm k})
\nonumber
\\
+&
{1 \over 2}
\sum_{\bm k}
{
|\Delta_{{\rm odd},p}|^2 \cos^2(\theta_{\bm k}) 
\over
\Lambda^2+|\Delta_{{\rm odd},p}|^2\cos^2(\theta_{\bm k})
}
\sum_{\alpha=\pm}
\nonumber
\\
&\times
\left(
1+\alpha{\xi_{\bm k} \over E_{{\rm odd},p}({\bm k})}
\right)
\delta(\omega-\alpha E_{{\rm odd},p}({\bm k})).
\label{eq.a17}
\end{align}
\par
When $\Lambda=0$, the first term in Eq. (\ref{eq.a17}) is absent, so that $\rho_{{\rm S},{\rm odd},p}(\omega)$ has the same form as SDOS in the even-frequency $p$-wave polar state. Indeed, as shown in Fig. \ref{fig2}(b), SDOS in this case has the V-shaped structure known in the polar state. In addition, the temperature dependence of $\Delta_{{\rm odd},p}(T)$ shown in Fig. \ref{fig1}(a) is also the same as that in the polar state. As a result, when $\Lambda=0$, we obtain the same temperature dependence of $(T_1T)^{-1}$ as in the even-frequency $p$-wave polar case. That is, $(T_1T)^{-1}\propto T^2$ at low temperatures\cite{Sigrist1991,Zheng2004} vanish at $T=0$, as seen in Fig. \ref{fig2}(a).
\par
When $\Lambda>0$, the first term in Eq. (\ref{eq.a17}) fills the V-shaped SDOS as shown in Fig. \ref{fig2}(b), which immediately explains the Korringa-law-like temperature dependence ($(T_1T)^{-1}={\rm const.}$)~\cite{Korringa1950} of the spin-lattice relaxation rate shown in Fig. \ref{fig2}(a). SDOS at $\omega=0$ is filled up to
\begin{equation}
\rho_{{\rm S},{\rm odd},p}(0)=\rho_{\rm N}(0)
{\Lambda \over |\Delta_{{\rm odd},p}|}
{\rm Tan}^{-1}
\left(
{|\Delta_{{\rm odd},p}| \over \Lambda}
\right).
\label{eq.SDOS0}
\end{equation}
Although Eq. (\ref{eq.SDOS0}) is always smaller than NDOS $\rho_{\rm N}(0)$, the inset in Fig. \ref{fig2}(b) shows that $\rho_{{\rm S},{\rm odd},p}(0)\simeq\rho_{\rm N}(0)$ for $\Lambda/|\Delta_{{\rm odd},p}|\gesim 1$. Thus, when $\Lambda/T_{\rm c}=3.4$ and 4 shown in Fig. \ref{fig2}(a), although the calculated $(T_1T)^{-1}$ only slightly decreases just below $T_{\rm c}$, it soon becomes constant with decreasing the temperature. Figure \ref{fig2}(a) also shows that this behavior well explains the recent experiment on CeRh$_{0.5}$Ir$_{0.5}$In$_5$~\cite{Kawasaki2020}. Although further analyses would be necessary to clarify the origin of the observed temperature dependence of $T_1^{-1}$ near QCP, our results indicate that the odd-frequency $p$-wave pairing state is really a promising candidate. 
\par
\begin{figure}[t]
\centering
\includegraphics[width=0.4\textwidth]{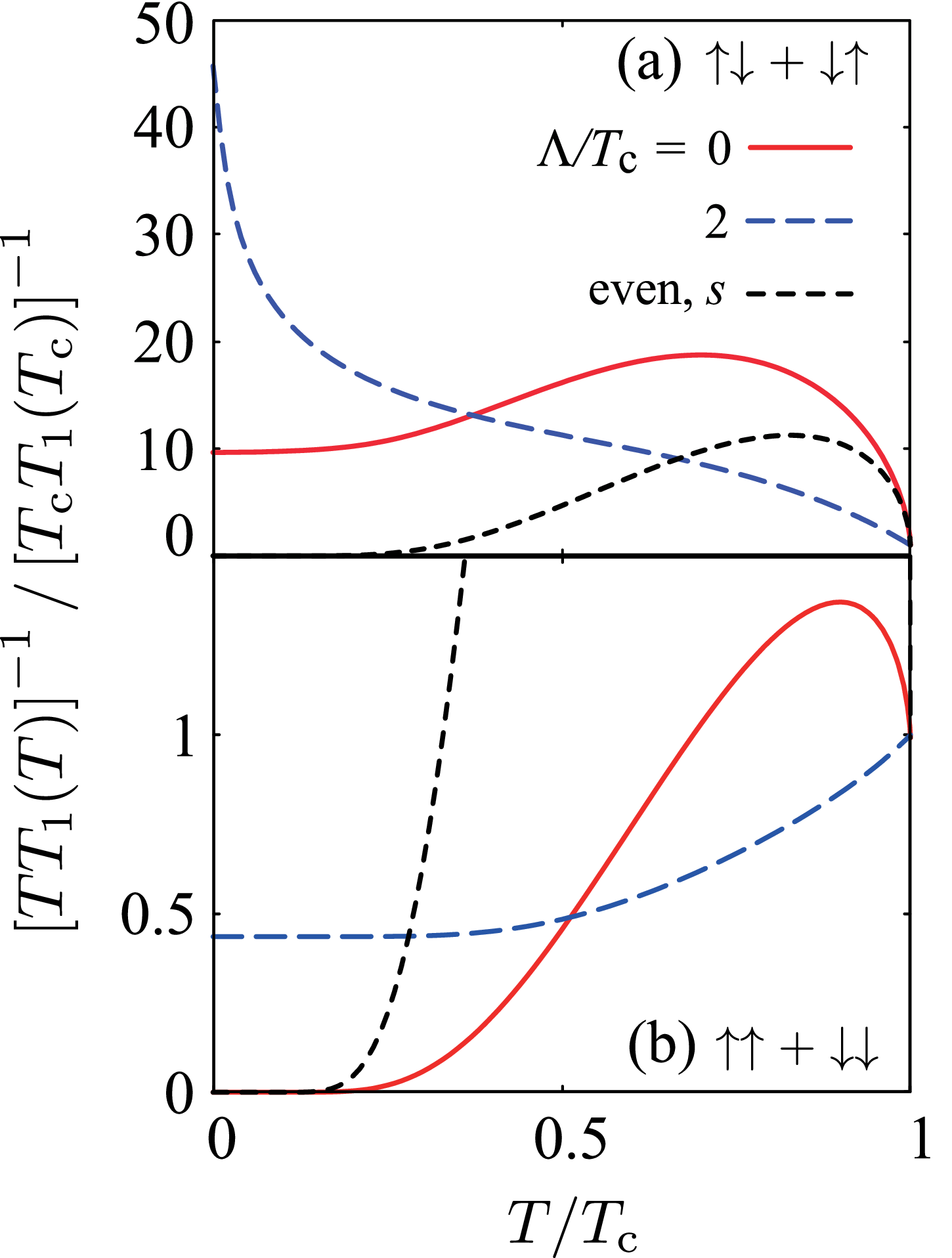}
\caption{Calculated $(T_1T)^{-1}$ as a function of the temperature in the odd-frequency spin-triplet $s$-wave state. (a) $\uparrow\downarrow+\downarrow\uparrow$ state. (b) $\uparrow\uparrow+\downarrow\downarrow$ state. For comparison, the even-frequency $s$-wave case is also shown as `even, $s$'. 
}
\label{fig3}
\end{figure}
\par
\par
\begin{figure}[t]
\centering
\includegraphics[width=0.4\textwidth]{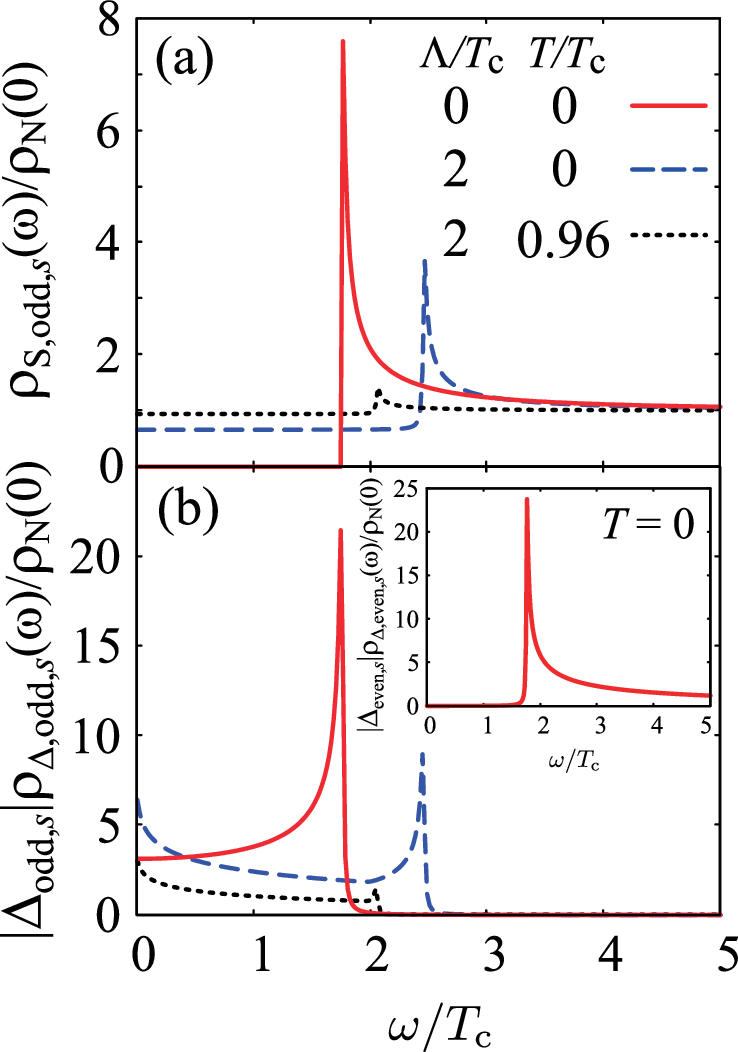}
\caption{Calculated (a) SDOS $\rho_{{\rm S},{\rm odd},s}(\omega)$ and (b) $\rho_{\Delta,{\rm odd},s}(\omega)$ in the odd-frequency $s$-wave state. In panel (a), $\rho_{{\rm S},{\rm odd},s}(\omega)=\rho_{{\rm S},{\rm even},s}(\omega)$ when $\Lambda=0$. The inset shows $\rho_{\Delta,{\rm even},s}(\omega)$ in the even-frequency $s$-wave BCS state at $T=0$.
}
\label{fig4}
\end{figure}
\par
\subsection{Odd-frequency $s$-wave state}
\par
Figure \ref{fig3} shows $(T_1T)^{-1}$ in the odd-frequency $s$-wave state. In the  $\uparrow \downarrow +\downarrow \uparrow$ case shown in panel (a), we find that, even when $\Lambda=0$, the behavior of $T_1^{-1}$ does not coincide with that in the conventional even-frequency $s$-wave BCS state. When $\Lambda=0$, because (1) $\Delta_{{\rm odd},s}(T)=\Delta_{{\rm even},s}(T)$ holds, and (2) SDOS in the odd-frequency $s$-wave state, given by
\begin{align}
\rho_{{\rm S},{\rm odd},s}(\omega)
=&
\sum_{\bm k}
{\Lambda^2 \over \Lambda^2+|\Delta_{{\rm odd},s}|^2}
\delta(\omega-\xi_{\bm k})
\nonumber
\\
+&
{1 \over 2}
\sum_{\bm k}
{
|\Delta_{{\rm odd},s}|^2
\over
\Lambda^2+|\Delta_{{\rm odd},s}|^2
}
\sum_{\alpha=\pm}
\nonumber
\\
&\times
\left(
1+\alpha{\xi_{\bm k} \over E_{{\rm odd},s}({\bm k})}
\right)
\delta(\omega-\alpha E_{{\rm odd},s}({\bm k})),
\label{eq.a16}
\end{align}
has the same form as that in the ordinary BCS state, given by
\begin{align}
\rho_{{\rm S},{\rm even},s}(\omega)
=&
{1 \over 2}
\sum_{\bm k}
\sum_{\alpha=\pm}
\nonumber
\\
&\times
\left(
1+\alpha{\xi_{\bm k} \over E_{{\rm even},s}({\bm k})}
\right)
\delta(\omega-\alpha E_{{\rm even},s}({\bm k})),
\label{eq.a15}
\end{align}
this difference is found to come from the coherence term (the second term in $[\cdot\cdot\cdot]$ in Eq. (\ref{eq.x1})). Indeed, evaluating Eqs. (\ref{eq.x4}) and (\ref{eq.x3}) when $\Lambda=0$, one obtains ($\omega\ge0$),
\begin{align}
\rho_{\Delta,{\rm odd},s}(\omega)
&=
\rho_{\rm N}(0)
{1 \over \sqrt{|\Delta_{{\rm odd},s}|^2-\omega^2}}
\Theta(|\Delta_{{\rm odd},s}|-\omega),
\label{eq.odds1}
\\
\rho_{\Delta,{\rm even},s}(\omega)
&=
\rho_{\rm N}(0)
{1 \over \sqrt{\omega^2-|\Delta_{{\rm odd},s}|^2}}
\Theta(\omega-|\Delta_{{\rm odd},s}|),
\label{eq.even1}
\end{align}
where $\Theta(x)$ is the step function. We then find that, in both states, although the peak structures at $\omega=|\Delta_l|$ in $\rho_{{\rm S},l}(\omega)$ and $\rho_{\Delta,l}(\omega)$ (see Fig. \ref{fig4}) produce the coherence peak just below $T_{\rm c}$, while $(T_1T)^{-1}$ in the even-frequency $s$-wave BCS state exponentially decreases at low temperatures due to the vanishing $\rho_{{\rm S},{\rm even},s}(\omega)$ and $\rho_{\Delta,{\rm even},s}(\omega)$ in the low energy region, $(T_1T)^{-1}$ remains non-zero down to $T=0$ in the odd-frequency $s$-wave case, because $\rho_{\Delta,{\rm odd},s}(\omega)>0$ around $\omega=0$ (see Fig. \ref{fig4}(b)).
\par
When $\Lambda>0$, the first term of SDOS $\rho_{{\rm S},{\rm odd},s}(\omega)$ in Eq. (\ref{eq.a16}) fills the BCS energy gap up to 
\begin{equation}
\rho_{{\rm S},{\rm odd},s}
\left(\omega\le\sqrt{\Lambda^2+|\Delta_{{\rm odd},s}|^2}\right)
=\rho_{\rm N}(0)
{\Lambda^2 \over \Lambda^2+|\Delta_{{\rm odd},s}|^2}.
\label{eq.sdos5}
\end{equation}
Thus, although the second term in Eq. (\ref{eq.a16}) still produces a peak at $\omega=\sqrt{\Lambda^2+|\Delta_{{\rm odd},s}|^2}$, the overall structure of SDOS becomes close to NDOS when $\Lambda/|\Delta_{{\rm odd},s}|\gesim 1$, as shown in Fig. \ref{fig4}(a). For $\rho_{\Delta,{\rm odd},s}(\omega)$, we obtain, from Eqs. (\ref{eq.21}), (\ref{eq.x4}), and (\ref{eq.xx4}),
\begin{align}
\rho_{\Delta,{\rm odd},s}(\omega)
=&
{\rho_{\rm N}(0) \over \pi}
\sum_{\bm k}
{\rm Re}
\left[
{1 \over \gamma_{{\rm odd},s}({\bm k},i\omega_n\to\omega+i\delta)}
\right.
\nonumber
\\
&\times
\left.
{1 \over (\omega+i\delta)^2-E_{{\rm odd},s}({\bm k})^2}
\right]
\nonumber
\\
\simeq&
{2 \over \pi}\rho_{\rm N}(0){\Lambda \over \Lambda^2+|\Delta_{{\rm odd},s}|^2}
\ln\left({\omega \over \Lambda}\right)~(\omega/\Lambda\ll 1).
\label{eq.app100}
\end{align}
In the last expression, we only retain the singular part around $\omega=0$. The last line in Eq. (\ref{eq.app100}) explains the low-energy behavior of $\rho_{\Delta,{\rm odd},s}(\omega)$ seen in Fig. \ref{fig4}(b) when $\Lambda/T_{\rm c}=2$\cite{notez}, as well as the enhancement of $(T_1T)^{-1}$ at low temperatures seen in Fig. \ref{fig3}. 
\par
In the case of the $\uparrow \uparrow + \downarrow \downarrow$ state shown in Fig. 3(b), while the SDOS term in Eq. (\ref{eq.x1}) is the same as that in the $\uparrow \downarrow +\downarrow \uparrow$ state shown in Fig 4(a), the coherence term vanishes identically (see Appendix B). As a result, $( T_{1} T)^{-1}$ in the $\uparrow \uparrow +\downarrow \downarrow$ state exhibits similar temperature dependence to that in the even-frequency $s$-wave state when $\Lambda =0$, and to that in the odd-frequency $p$-wave state when $\Lambda /T_{\mathrm{c}} =2.$ In this sence, $( T_{1} T)^{-1}$ in the odd-frequency spin-triplet $s$-wave state is sensitive to the detailed spin structure of the Cooper pair.
\par
\par
\section{Summary}
\par
To summarize, we have discussed the nuclear spin-lattice relaxation rate $T_1^{-1}$ in the odd-frequency $s$-wave and $p$-wave superconducting states. Using the model odd-frequency pairing interaction which satisfies the positivity of SDOS~\cite{Iwasaki2024}, and employing the combined path-integral formalism with the saddle-point approximation which gives the stable superconducting state below $T_{\rm c}$~\cite{Solenov2009,Kusunose2011B}, we examined how the odd-frequency pairing affects the temperature dependence of $T_1^{-1}$.
\par
In the odd-frequency $p$-wave superconductivity, we showed that, under a certain condition, SDOS around $\omega=0$ is filled up, leading to the Korringa-law-like $T$-linear behavior of $T_1^{-1}$ below $T_{\rm c}$, even in the clean system without non-magnetic impurities. We also showed that this result well explains the recent experiment on CeRh$_{0.5}$Ir$_{0.5}$In$_5$~\cite{Kawasaki2020}, where the possibility of the odd-frequency $p$-wave state has recently been discussed near QCP.
\par 
In the $s$-wave case, the behavior of $T_1^{-1}$ depends on the spin structure of the superconducting order parameter. In the $\uparrow\uparrow+\downarrow\downarrow $ state, the coherence term does not contribute to $T_1^{-1}$, leading to the similar behavior of $T_1^{-1}$ to the odd-frequency $p$-wave case. In contrast, the coherence term contributes to $T_1^{-1}$ in the $\uparrow \downarrow +\downarrow \uparrow $ state, which brings about the anomalous enhancement of $(T_1T)^{-1}$ far below $T_{\rm c}$.
\par
We finally summarize remaining problems to be solved: (1) Although our model odd-frequency $p$-wave state can explain the recent experiment on CeRh$_{0.5}$Ir$_{0.5}$In$_5$, it is still unclear whether or not this scenario is superior to the `nonmagnetic impurity scenario.' To confirm this, it would be effective to explore a phenomenon where the two scenarios give qualitatively different results. (2) We did not discuss the pairing mechanism in this paper, but simply assumed a model odd-frequency attractive interaction, Thus, it is also a crucial issue to check the universality of our results. For example, the $T$-linear behavior of $T_1^{-1}$ would also be obtained, even when the polar-type pairing interaction assumed in this paper is replaced by other $p$-wave type momentum dependence, because the non-zero SDOS around $\omega=0$ coming from the odd-frequency dependence of the order parameter plays an essential role in obtaining this result. On the other hand, because the logarithmic singularity of $\rho_{\Delta,{\rm odd},s}(\omega)$ at $\omega=0$ shown in Fig, \ref{fig4} originates from the detailed $\omega$-dependence of the basis function $\gamma_{{\rm odd},s}({\bm k},i\omega_n\to\omega+i\delta)$, the resulting enhancement of $(T_1T)^{-1}$ at low temperatures in the odd-frequency $s$-wave case (see Fig. \ref{fig3}) might be model dependent. In the current stage of research, there are very few model odd-frequency interaction that guarantees the positivity of SDOS. Thus, in order to check the universality of our results, it is important to explore other odd-frequency models which also satisfy this required condition. (3) Our model does not take into account the details of CeRh$_{0.5}$Ir$_{0.5}$In$_5$, so that it also remains as our future problem to see how the results obtained in this paper are altered in a more realistic model.
\par 
In addition to these, we also note that we have treated the odd-frequency superconducting state within the mean-field level. Regarding this, the BCS-BEC crossover phenomenon\cite{Randeria1995} of odd-frequency $s$-wave superconductivity was recently discussed in a model transition metal oxide\cite{Inokuma2024}. Thus, it is also an interesting challenge to extend our theory to include effects of pairing fluctuations beyond the mean-field level. Since odd-frequency pairing states have also been discussed in cold Fermi gas physics, this extension would contribute to the further development of this field. At present, although the bulk odd-frequency superconductivity has not experimentally been confirmed yet, since CeRh$_{0.5}$Ir$_{0.5}$In$_5$ is considered as a promising candidate, our results would contribute to the research toward the realization of this unique pairing state in the bulk system.
\par
\par
\section*{Acknowledgments}
\begin{acknowledgments}
We thank S. Kawasaki for providing us with his experimental data. S.I. also thanks Y. Tanaka for providing him with useful information about odd-frequency superconductivity. S.I. was supported by a Grant-in-Aid for JST SPRING (Grant No.JPMJSP2123). Y.O. was supported by a Grant-in-Aid for Scientific Research from MEXT and JSPS in Japan (Grant No.JP22K03486).
\end{acknowledgments}
\par
\appendix
\par
\section{Derivation of Eq.(\ref{eq.x1})}
\par
In this appendix, we derive Eq. (\ref{eq.x1}). Evaluating the statistical average $\langle\cdot\cdot\cdot\rangle$ in Eq. (\ref{eq.a2}) in the mean-field level, one can decompose it into the sum of the product of the mean-field single-particle thermal Green's function in Eq. (\ref{eq.21}), by using the Wick's theorem\cite{Stoof2009}. The results is
\begin{align}
&\sum _{k,k'} \langle \overline{\psi }_{k,\downarrow} \psi _{k+q,\uparrow}\overline{\psi }_{k'+q,\uparrow} \psi _{k',\downarrow} \rangle  
\nonumber
\\
&=
\sum _{k,k'}
\left[ 
\langle \psi_{k'+q,\uparrow} \overline{\psi }_{k+q,\uparrow}\rangle
\langle \overline{\psi }_{k,\downarrow} \psi _{k',\downarrow} \rangle 
\right.
\nonumber
\\
&~~~~~~~~~~~~~~~~~~~~~~~~~~~~~~~-
\left.
\langle \psi _{k+q,\uparrow} \psi _{k',\downarrow} \rangle 
\langle \overline{\psi }_{k,\downarrow}\overline{\psi }_{k'+q,\uparrow} \rangle 
\right]
\nonumber
\\
& =
\sum _{k}\left[ G_{l}^{( 1,1)}( q-k) G_{l}^{( 2,2)}( k) -G_{l}^{( 1,2)}( k) G_{l}^{( 2,1)}( q-k) \right].
\label{eq.a9}
\end{align}
In obtaining the last expression in Eq.~(\ref{eq.a9}), we have used the following relation:
\begin{align}
{\hat G}_l(k)
&=
\begin{pmatrix}
G_l^{(1,1)}(k)
& 
G_l^{(1,2)}(k) 
\\
G_l^{(2,1)}(k)  
& 
G_l^{(2,2)}(k) 
\end{pmatrix} 
\nonumber
\\
&=
\begin{pmatrix}
-\langle \psi _{k,\uparrow}\overline{\psi }_{k,\uparrow} \rangle  
& 
-\langle \psi _{k,\uparrow} \psi _{-k,\downarrow} \rangle 
\\
-\langle \overline{\psi }_{-k,\downarrow}\overline{\psi }_{k,\uparrow} \rangle  
& 
-\langle \overline{\psi }_{-k,\downarrow} \psi _{-k,\downarrow} \rangle 
\end{pmatrix} 
\label{eq.a8}
\end{align}
\par
We next introduce the spectral representation of the Green's function:~\cite{Altrand}
\begin{equation}
\hat{G}_{l}({\bm k} ,i\omega_n) 
=
\int _{-\infty }^{\infty } d\omega \frac{\hat{A}_{l}({\bm k} ,\omega )}{i\omega _{n} -\omega },
\label{eq.a10}
\end{equation}
where 
\begin{align}
\hat{A}_{l}({\bm k} ,\omega ) 
&=-\frac{1}{2\pi i}\left[
\hat{G}_{l}\left({\bm k} ,i\omega _{n}\rightarrow \omega +i\delta \right)\Bigl|_{\omega _{n}  >0} 
\right.
\nonumber
\\
&~~~~~~~\left.
-\hat{G}_{l}\left({\bm k} ,i\omega _{n}\rightarrow \omega -i\delta \right)\Bigl|_{\omega _{n} < 0}
\right]
\label{eq.a11}
\end{align}
is the spectral weight. Substituting Eqs. (\ref{eq.a9}) and (\ref{eq.a10}) into Eq. (\ref{eq.a2}), we obtain
\begin{align}
&\chi _{l}({\bm q},i\nu_{m}) 
\nonumber
\\
&=
T\sum _{{\bm k},i\omega_n}
\int_{-\infty }^{\infty } d\omega 
\int_{-\infty }^{\infty } d\omega'
\frac{1}{[i\nu _{m} -i\omega _n-\omega '][i\omega_n-\omega ]}
\nonumber
\\
&\times
\Big[
A_{l}^{(1,1)}({\bm q}-{\bm k},\omega') 
A_{l}^{(2,2)}({\bm k},\omega) 
\nonumber
\\
&~~~~~~~~~~~~~~~~~~- 
A_{l}^{(1,2)}({\bm k},\omega) 
A_{l}^{(2,1)}({\bm q}-{\bm k},\omega') 
\Big].
\label{eq.aa1}
\end{align}
Carrying out the $\omega_n$-summation in Eq.~(\ref{eq.aa1}), one has, after taking the analytic continuation ($i\nu_m\to\nu+i\delta$), 
\begin{align}
&\chi_{l}\left(\mathbf{q} ,i\nu _{m}\rightarrow \nu +i\delta \right)
=
\int_{-\infty }^{\infty } d\omega 
\int_{-\infty }^{\infty } d\omega '
\frac{f(\omega)-f(-\omega')}{\nu+i\delta-\omega-\omega'}
\nonumber
\\
&\times
\sum_{\bm k} 
\Big[
A_{l}^{(1,1)}({\bm q}-{\bm k},\omega') 
A_{l}^{(2,2)}({\bm k},\omega) 
\nonumber
\\
&~~~~~~~~~~~~~~~~~~- 
A_{l}^{(1,2)}({\bm k},\omega) 
A_{l}^{(2,1)}({\bm q}-{\bm k},\omega') 
\Big].
\label{eq.app3}
\end{align}
Here, $f(\omega)=[\exp(\beta\omega)+1]^{-1}$ is the Fermi distribution function. Substituting Eq. (\ref{eq.app3}) into Eq. (\ref{eq.a1}), one obtains
\begin{align}
&\frac
{\left[TT_{1,l}( T)\right]^{-1}}
{\left[T_{\rm c}T_{1,l}(T_{\rm c})\right]^{-1}}
=
{\beta \over4\rho_{\rm N}(0)^2}
\int_{-\infty }^{\infty } d\omega 
{\rm sech}^2
\left(
{\beta\omega \over 2}
\right)
\nonumber
\\
&\times
\sum _{{\bm k} ,{\bm k} '}
\Big[
A_{l}^{(1,1)}({\bm k} ',\omega ) 
A_{l}^{( 2,2)}({\bm k} ,-\omega ) 
\nonumber
\\
&~~~~~~~~~~~~~~~~~~~~~~~~~- 
A_{l}^{(1,2)}({\bm k} ,\omega ) 
A_{l}^{(2,1)}({\bm k} ',-\omega )
\Big],
\label{eq.a14}
\end{align}
where 
$\left[T_{\rm c}T_{1,l}(T_{\rm c})\right]^{-1}=\pi C \rho_{\rm N}(0)^2$~\cite{Korringa1950}.
\par
The diagonal components $G^{i,i}(k)~(i=1,2)$ of the Nambu thermal Green's function in Eq. (\ref{eq.21}) have the following symmetry properties:
\begin{align}
G_{l}^{( 2,2)}&({\bm k} ,i\omega _{n})
=-G_{l}^{( 1,1)}({\bm k} ,-i\omega _{n}),
\label{eq.a18}
\\
G_{l}^{( 1,1)}&\left({\bm k} ,i\omega _{n}\rightarrow \omega -i\delta \right)\Bigl|_{\omega _{n} < 0} 
\nonumber
\\
&=G_{l}^{( 1,1)}\left({\bm k} ,i\omega _{n}\rightarrow \omega +i\delta \right)\Bigl|_{\omega _{n}  >0}^{*}.
\label{eq.a19}
\end{align}
Using these, we find that the spectral weights $A_{l}^{(i,i)}({\bm k},\omega)~(i=1,2)$ are related to each other as, 
\begin{align}
A_{l}^{(1,1)}({\bm k},\omega) 
&=
A_{l}^{(2,2)}({\bm k},-\omega)
\nonumber
\\
&=
\left. -\frac{1}{\pi }\mathrm{Im} G_{l}^{(1,1)}\left({\bm k} ,i\omega _{n}\rightarrow \omega +i\delta \right)\right|_{\omega _{n}  >0},
\label{eq.a20}
\end{align}
which gives
\begin{equation}
\sum_{{\bm k},{\bm k}'}
A_{l}^{(1,1)}({\bm k}',\omega)A_{l}^{(2,2)}({\bm k},-\omega)=
\rho_{{\rm S},l}(\omega)^2,
\label{eq.ap200}
\end{equation}
where $\rho_{{\rm S},l}(\omega)$ is given in Eq. (\ref{eq.x2}).
\par
For $A_{l}^{(1,2)}({\bm k}',\omega)$ and $A_{l}^{(2,1)}({\bm k}',\omega)$ in the odd-frequency $s$-wave case, we note that $F_{{\rm odd},s}$ given in Eq. (\ref{eq.xx4}) has the following symmetry properties,
\begin{align}
F_{{\rm odd},s}&({\bm k},i\omega_n\rightarrow -\omega +i\delta)
\Bigl|_{\omega_n>0} 
\nonumber
\\
&=
-F_{{\rm odd},s}({\bm k},i\omega_n\rightarrow \omega -i\delta)
\Bigl|_{\omega_n<0},
\label{eq.a21}
\\
F_{{\rm odd},s}&({\bm k},i\omega_n\rightarrow -\omega -i\delta)
\Bigl|_{\omega_n<0} 
\nonumber
\\
& =
-F_{{\rm odd},s}({\bm k},i\omega_n\rightarrow \omega +i\delta)
\Bigl|_{\omega_n>0},
\label{eq.a22}
\\
F_{{\rm odd},s}&({\bm k},i\omega_n\rightarrow \omega -i\delta)
\Bigl|_{\omega_n<0} 
\nonumber
\\
& =
-F_{{\rm odd},s}({\bm k},i\omega_n\rightarrow \omega +i\delta)
\Bigl|_{\omega_n>0}^{*}.
\label{eq.a23}
\end{align}
Then we obtain
\begin{align}
&A_{{\rm odd},s}^{(1,2)}({\bm k},\omega) 
\nonumber
\\
&=-{\Delta_{{\rm odd},s} \over \pi i}
{\rm Re}
\left[F_{{\rm odd},s}({\rm k} ,i\omega _{n}\rightarrow \omega +i\delta)
\right]
\Bigl|_{\omega_n>0},
\label{eq.a24}
\\
&A_{{\rm odd},s}^{(2,1)}({\bm k},\omega) 
\nonumber
\\
&=-{\Delta_{{\rm odd},s}^* \over \pi i}
{\rm Re}
\left[F_{{\rm odd},s}({\bm k} ,i\omega _{n}\rightarrow \omega +i\delta)
\right]
\Bigl|_{\omega_n>0}.
\label{eq.a24b}
\end{align}
\par
The same symmetry properties as Eqs. (\ref{eq.a21})-(\ref{eq.a23}) are also satisfied in the odd-frequency $p$-wave case, leading to
\begin{align}
&A_{{\rm odd},p}^{(1,2)}({\bm k},\omega) 
\nonumber
\\
&=-{\Delta_{{\rm odd},p} \over \pi i}
{\rm Re}
\left[F_{{\rm odd},p}({\bm k} ,i\omega _{n}\rightarrow \omega +i\delta)
\right]
\Bigl|_{\omega_n>0},
\label{eq.a25}
\end{align}
\begin{align}
&A_{{\rm odd},p}^{(2,1)}({\bm k},\omega) 
\nonumber
\\
&=-{\Delta_{{\rm odd},p}^* \over \pi i}
{\rm Re}
\left[F_{{\rm odd},p}({\bm k} ,i\omega _{n}\rightarrow \omega +i\delta)
\right]
\Bigl|_{\omega_n>0}.
\label{eq.a25b}
\end{align}
As a result, for the odd-frequency $p$-wave and $s$-wave states, one has
\begin{equation}
\sum_{{\bm k},{\bm k}'}
A_{l}^{(1,2)}({\bm k}',\omega)A_{l}^{(2,1)}({\bm k},-\omega)=
-|\Delta_l|^2\rho_{\Delta,l}(\omega)^2,
\label{eq.ap300}
\end{equation}
where $\rho_{\Delta,{\rm odd},s}(\omega)$ and $\rho_{\Delta,{\rm odd},p}$ are given in Eqs. (\ref{eq.x4}) and (\ref{eq.x5}), respectively.
\par
The even-frequency $s$-wave has the following symmetry properties:
\begin{align}
F_{{\rm even},s}&({\bm k},i\omega_n\rightarrow -\omega +i\delta)
\Bigl|_{\omega_n>0} 
\nonumber
\\
& =+F_{{\rm even},s}({\bm k},i\omega_n\rightarrow \omega -i\delta)
\Bigl|_{\omega_n<0},
\label{eq.a26}
\\
F_{{\rm even},s}&({\bm k},i\omega_n\rightarrow -\omega -i\delta)
\Bigl|_{\omega_n<0} 
\nonumber
\\
& =+F_{{\rm even},s}({\rm k},i\omega_n\rightarrow \omega +i\delta)
\Bigl|_{\omega_n>0} ,
\label{eq.a27}
\\
F_{{\rm even},s}&({\bm k},i\omega_n\rightarrow \omega -i\delta)
\Bigl|_{\omega_n<0} 
\nonumber
\\
& =+F_{{\rm even},s}({\bm k},i\omega_n\rightarrow \omega +i\delta)
\Bigl|_{\omega_n>0}^{*} .
\label{eq.a28}
\end{align}
Using Eqs. (\ref{eq.a11}), and (\ref{eq.a26})-(\ref{eq.a28}), we obtain
\begin{align}
&A_{{\rm even},s}^{( 1,2)}({\bm k},\omega)
\nonumber
\\
&=-{\Delta_{{\rm even},s} \over \pi}
\sum_{\bm k}{\rm Im}
\left[
F_{{\rm even},s}({\bm k},i\omega_n\rightarrow \omega +i\delta)
\right]
\Bigl|_{\omega_n>0},
\label{eq.a29}
\\
&A_{{\rm even},s}^{(2,1)}({\bm k},\omega)
\nonumber
\\
&={\Delta_{{\rm even},s}^* \over \pi}
\sum_{\bm k}{\rm Im}
\left[
F_{{\rm even},s}({\bm k},i\omega_n\rightarrow \omega +i\delta)
\right]
\Bigl|_{\omega_n>0},
\label{eq.a29b}
\end{align}
which gives
\begin{align}
\sum_{{\bm k},{\bm k}'}
A_{{\rm even},s}^{(1,2)}({\bm k}',\omega)
&A_{{\rm even},s}^{(2,1)}({\bm k},-\omega)
\nonumber
\\
&=
-|\Delta_{{\rm even},s}|^2\rho_{\Delta,{\rm even},s}(\omega)^2,
\label{eq.ap100}
\end{align}
where $\rho_{\Delta,{\rm even},s}(\omega)$ is given in Eq. (\ref{eq.x3}).
\par
From Eqs. (\ref{eq.a14}), (\ref{eq.ap200}), (\ref{eq.ap300}), and (\ref{eq.ap100}), we obtain Eq. (\ref{eq.x1}).
\par
\section{$T_1^{-1}$ in the odd-frequency $\uparrow \uparrow+\downarrow\downarrow$ $s$-wave state}
\par
To derive $T_{1}^{-1}$ in the odd-frequency $\uparrow\uparrow +\downarrow\downarrow$ $s$-wave state, we start from the path-integral representation of partition function $Z$ in Eq. (\ref{eq.1}) where the interaction part $S_1$ of the action is replaced by \cite{Iwasaki2024,Kusunose2011A,Kusunose2011B}
\begin{align}
S_{1} =&
-\frac{U_{\mathrm{odd,s}}}{2\beta }\sum _{q,k,k'} \gamma {\textstyle _{\mathrm{odd} ,s}\left(k+\frac{q}{2}\right) \gamma _{\mathrm{odd} ,s}\left(k'+\frac{q}{2}\right)}
\nonumber
\\
&\times\overline{\psi }_{k+q,\beta }\overline{\psi }_{-k,\alpha } \psi _{-k',\alpha } \psi _{k'+q,\beta }.
\end{align}
Here, the subscripts $\alpha,\beta =\uparrow ,\downarrow $ denote electron spin. As in Sec. 2.1, introducing complex field $\Phi _{\alpha \beta }$ and its complex conjugate field $\overline{\Phi }_{\alpha \beta }$ describing Cooper pair bose field by Hubbard-Stratonovich transformation, and employing saddle-point approximation with saddle point as $\Phi ^{\alpha \beta } =\beta \Delta _{\alpha \beta } \delta _{q,0} ,\overline{\Phi }^{\alpha \beta } =\beta \Delta _{\alpha \beta }^{*} \delta _{q,0}$ \cite{Kusunose2011B}, we obtain the saddle-point partition function $Z_{\mathrm{sp}}$ in Eq. (\ref{eq.b1}) where the saddle-point action $S_{\mathrm{sp}}$ in Eq. (\ref{eq.b2}) is replaced by \cite{Kusunose2011B} 
\begin{align}
S_{1} 
&=\frac{1}{2}
\sum _{k,\alpha ,\beta }
\begin{pmatrix}
\overline{\psi }_{k,\beta } & \psi _{-k,\alpha }
\end{pmatrix}
\nonumber
\\
&\times
\begin{pmatrix}
-i\omega _{n} +\xi _{{\bm k}} & \Delta _{\alpha \beta } \gamma ({\bm k} ,i\omega _{n})\\
\Delta _{\alpha \beta }^{*} \gamma ({\bm k} ,i\omega _{n}) & -i\omega _{n} -\xi _{{\bm k}}
\end{pmatrix}
\begin{pmatrix}
\psi _{k,\beta }\\
\overline{\psi }_{-k,\alpha }
\end{pmatrix}
\nonumber
\\
&+\beta \sum _{{\bm k}} \xi _{{\bm k}} 
+\sum _{\alpha ,\beta }\frac{\beta|\Delta _{\alpha ,\beta } |^{2}}{2U_{\mathrm{odd,s}}} .
\label{eq.bb2}
\end{align}
To describe the $\uparrow \uparrow + \downarrow \downarrow$ state, we set $\Delta _{\alpha \beta } =\Delta _{\mathrm{odd} .s} \delta _{\alpha \beta }$ \cite{Kusunose2011A} (In Sec. 2.1, we have chosen $\Delta _{\alpha \beta } =\Delta _{\mathrm{odd,s}}\left[\hat{\sigma }_{1}\right]_{\alpha \beta }$ with $\sigma _{1}$ being the Pauli matrix acting on spin space). In this case, one can rewrite the saddle-point action $S_{\mathrm{SP}}$ in Eq. (\ref{eq.bb2}) as 
\begin{align}
S_{\mathrm{sp}} 
=
\frac{1}{2}\sum _{k,\alpha } \Psi _{k,\alpha }^{\dagger }&\left[ -\hat{G}_{\mathrm{odd} ,s}^{-1}( k)\right]\Psi _{k,\alpha }
\nonumber
\\
&+\beta \sum _{{\bm k}} \xi _{{\bm k}} +\frac{\beta |\Delta _{\mathrm{odd} ,s} |^{2}}{U_{\mathrm{odd} ,s}} ,
\end{align}
where $\Psi _{\alpha } =( \psi _{k,\alpha } ,\bar{\psi}_{-k,\alpha })^{\mathrm{T}}$, and $\hat{G}_{\mathrm{odd} ,s}( k)$ is given in Eq. (\ref{eq.21}). Executing the fermion path-integrals in $Z_{\mathrm{sp}}$ in Eq. (\ref{eq.b1}), one again obtains the thermodynamic potential $\Omega_{\mathrm{MF}}$ in Eq. (\ref{eq.24}). Thus, the superconducting order parameter, which is determined from the saddle point condition in Eq (\ref{eq.23}), is also the same as that in the $\uparrow \downarrow +\downarrow \uparrow $ state.
\par
In the present $\uparrow \uparrow + \downarrow \downarrow$ state, the statistical average in Eq. (\ref{eq.a2}) is evaluated as, within the mean-field scheme, 
\begin{align}
\sum _{k,k'} \langle \overline{\psi }_{\downarrow k} \psi _{\uparrow k+q}\overline{\psi }_{\uparrow k'+q} \psi _{\downarrow k'} \rangle  & =\sum _{k,k'} \langle \overline{\psi }_{\downarrow k} \psi _{\downarrow k'} \rangle \langle \overline{\psi }_{\uparrow k'+q} \psi _{\uparrow k+q} \rangle 
\nonumber\\
& =\sum _{k} G_{\mathrm{odd} ,s}^{( 2,2)}( k) G_{\mathrm{odd} ,s}^{( 1,1)}( q-k) ,
\label{eq.bb4}
\end{align}
where we have used the following relation in obtaining the last line in Eq. (\ref{eq.bb4}):
\begin{align}
\hat{G}_{\mathrm{odd} ,s}( k) 
&=
\begin{pmatrix}
G_{\mathrm{odd,} s}^{( 1,1)}( k) & G_{\mathrm{odd,} s}^{( 1,2)}( k)\\
G_{\mathrm{odd,} s}^{( 2,1)}( k) & G_{\mathrm{odd,} s}^{( 2,2)}( k)
\end{pmatrix} 
\nonumber
\\
&=
\begin{pmatrix}
-\langle \psi _{k,\alpha }\overline{\psi }_{k,\alpha } \rangle  & -\langle \psi _{k,\alpha } \psi _{-k,\alpha } \rangle \\
-\langle \overline{\psi }_{-k,\alpha }\overline{\psi }_{k,\alpha } \rangle  & -\langle \overline{\psi }_{-k,\alpha } \psi _{-k,\alpha } \rangle 
\end{pmatrix} .
\label{eq.bb5}
\end{align}
Substituting the explicit forms of $G_{\mathrm{odd} ,s}^{( 2,2)}( k)$ and $G_{\mathrm{odd} ,s}^{( 1,1)}( q-k)$ into Eq. (\ref{eq.bb4}), and carrying out the Matsubara frequency summation, we obtain 
\begin{equation}
\frac{\left[ TT_{1}( T)\right]^{-1}}{\left[ T_{\mathrm{c}} T_{1}( T_{\mathrm{c}})\right]^{-1}} =\frac{\beta }{4\rho _{\mathrm{N}}^{2}( 0)}\int _{-\infty }^{\infty } d \omega \rho _{\mathrm{S,odd} ,s}^{2}( \omega )\sech^{2}\left(\frac{\beta \omega }{2}\right).
\label{eq.bb6}
\end{equation}
Equation (\ref{eq.bb6}) corresponds to Eq. (\ref{eq.x1}) where the second term in $[\cdots]$ is absent.
\par


\begin{thebibliography}{99}
\bibitem{Linder2019} J. Linder and A. V. Balatsky, Rev. Mod. Phys. \textbf{91}, 045005 (2019).
\bibitem{Berezinskii1974} V. L. Berezinskii, JETP Lett. \textbf{20}, 287 (1974).
\bibitem{Emery1992} V. J. Emery and S. Kivelson, Phys. Rev. B \textbf{46}, 10812 (1992).
\bibitem{Tsvelik1993} P. Coleman, E. Miranda, and A. Tsvelik, Phys. Rev. Lett. \textbf{70}, 2960 (1993).
\bibitem{Tsvelik1994} P. Coleman, E. Miranda, and A. Tsvelik, Phys. Rev. B \textbf{49}, 8955 (1994).
\bibitem{Fuseya2003} Y. Fuseya, H. Kohno, and K. Miyake, J. Phys. Soc. Jpn. \textbf{72}, 2914 (2003).
\bibitem{Shigeta2012} K. Shigeta, S. Onari, and Y. Tanaka, Phys. Rev. B \textbf{85}, 224509 (2012).
\bibitem{Hoshino2014A} S. Hoshino and Y. Kuramoto, Phys. Rev. Lett. \textbf{112}, 167204 (2014).
\bibitem{Hoshino2014B} S. Hoshino, Phys. Rev. B \textbf{90}, 115154 (2014).
\bibitem{Hoshino2016} S. Hoshino, K. Yada, and Y. Tanaka, Phys. Rev. B \textbf{93}, 224511(2016).
\bibitem{Funaki2014} H. Funaki and H. Shimahara, J. Phys. Soc. Jpn. \textbf{83}, 123704 (2014).
\bibitem{Otsuki2015} J. Otsuki, Phys. Rev. Lett. \textbf{115}, 036404 (2015).
\bibitem{Shigeta2009} K. Shigeta, S. Onari, K. Yada, and Y. Tanaka, Phys. Rev. B \textbf{79}, 174507 (2009).
\bibitem{Shigeta2011} K. Shigeta, Y. Tanaka, K. Kuroki, S. Onari, and H. Aizawa, Phys. Rev. B \textbf{83}, 140509(R) (2011).
\bibitem{Pratt2013} F. L. Pratt, T. Lancaster, S. J. Blundell, and C. Baines, Phys. Rev. Lett. \textbf{110}, 107005 (2013)
107005.
\bibitem{Fukui2018} K. Fukui and Y. Kato, J. Phys. Soc. Jpn. \textbf{87}, 014706 (2018).
\bibitem{Fukui2019} K. Fukui and Y. Kato, J. Low. Temp. Phys. \textbf{196}, 234 (2019).
\bibitem{Yoshida2021} S. Yoshida, K. Yada, and Y. Tanaka, Phys. Rev. B \textbf{104}, 094506 (2021).
\bibitem{Balatsky1992} A. Balatsky and E. Abrahams, Phys. Rev. B \textbf{45}, 13125  (1992).
\bibitem{Abrahams1993} E. Abrahams, A. Balatsky, J. R. Schrieffer, and P. B. Allen, Phys. Rev. B \textbf{47}, 513 (1993).
\bibitem{Matsubara2021} S. Matsubara, Y. Tanaka, and H. Kontani, Phys. Rev. B \textbf{103}, 245138 (2021).
\bibitem{Inokuma2024} Y. Inokuma and Y. Ono, J. Phys. Soc. Jpn \textbf{93}, 043701 (2024).
\bibitem{Kalas2008} R. M. Kalas, A. V. Balatsky, and D. Mozyrsky, Phys. Rev. B \textbf{78}, 184513 (2008).
\bibitem{Arzamasovs2018} M. Arzamasovs and B. Liu, Phys. Rev. A \textbf{97}, 043607 (2018).
\bibitem{Iwasaki2024} S. Iwasaki, T. Kawamura, K. Manabe, and Y. Ohashi, Phys. Rev. A \textbf{109}, 063309 (2024).
\bibitem{Kusunose2011A} H. Kusunose, Y. Fuseya, and K. Miyake, J. Phys. Soc. Jpn. \textbf{80}, 044711 (2011).
\bibitem{Matsumoto2012} M. Matsumoto, M. Koga, and H. Kusunose, J. Phys. Soc. Jpn. \textbf{81}, 033702 (2012).
\bibitem{Kusunose2012} H. Kusunose, M. Matsumoto, and M. Koga, Phys. Rev. B \textbf{85}, 174528 (2012).
\bibitem{Kawasaki2020} S. Kawasaki, T. Oka, A. Sorime, Y. Kogame, K. Uemoto, K. Matano, J. Guo, S. Cai, L. Sun, J. L. Sarrao, J. D. Thompson and G.-q. Zheng, Commun. Phys. \textbf{3}, 148 (2020).
\bibitem{Parks} D. M. Ginsberg, and L. C. Hebel in \textit{Superconductivity}, Vols. I, ed. by R. D. Parks (Dekker, New York, 1969) Chap. 4.
\bibitem{Hebel1960} L. C. Hebel and C. P. Slichter Phys. Rev. \textbf{107}, 901 (1957).
\bibitem{Moriya1956} T. Moriya, Prog. Theor. Phys. \textbf{16}, 23 (1956).
\bibitem{Moriya1963} T. Moriya, J. Phys. Soc. Jpn. \textbf{18}, 516 (1963).
\bibitem{Pfleiderer2008} C. Pfleiderer, Rev. Mod. Phys. \textbf{81}, 1551 (2008).
\bibitem{Sigrist1991} M. Sigrist and K. Ueda, Rev. Mod. Phys. \textbf{63}, 239 (1991).
\bibitem{Bang2004} Y. Bang, M. J. Graf, A. V. Balatsky, and J. D. Thompson, Phys. Rev. B \textbf{69}, 014505 (2004).
\bibitem{Korringa1950} J. Korringa, Physica \textbf{16}, 601 (1950).
\bibitem{note2} Although the pairing mechanism of CeRh$_{0.5}$Ir$_{0.5}$In$_{5}$ is still unclear, the simplest (even-frequency) spin-singlet $s$-wave BCS state seems difficult to realize due to the strong electron correlation. Thus, if strong antiferromagnetic spin fluctuations near QCP enhance a spin-singlet pairing interaction, the even-frequency $d$-wave state and the odd-frequency $p$-wave state are candidates for spin-singlet superconductivity.
\bibitem{Kawasaki2003} S. Kawasaki, T. Mito, Y. Kawasaki, G.-q. Zheng, Y. Kitaoka, D. Aoki, Y. Haga, and Y. Onuki, Phys. Rev. Lett. \textbf{91}, 137001 (2003).
\bibitem{Zheng2004} Guo-qing Zheng, N. Yamaguchi, H. Kan, Y. Kitaoka, J. L. Sarrao, P. G. Pagliuso, N. O. Moreno, and J. D. Thompson, Phys. Rev. B \textbf{70}, 014511 (2004).
\bibitem{Schmitt-Rink1986} Schmitt-Rink, K. Miyake, and C. M. Varma, Phys. Rev. Lett. \textbf{57}, 2575, (1986).
\bibitem{Ishida1993} K. Ishida, Y. Kitaoka, N. Ogata, T. Kamino, K. Asayama, J. R. Cooper, and N. Athanassopoulou, J. Phys. Soc. Jpn. \textbf{62}, 2803 (1993).
\bibitem{Ishida2000} K. Ishida, H. Mukuda, Y. Kitaoka, Z. Q. Mao, Y. Mori, and Y. Maeno, Phys. Rev. Lett. \textbf{84}, 5387 (2000).
\bibitem{Hotta2000} T. Hotta, J. Phys. Soc. Jpn. \textbf{62}, 274 (1993).
\bibitem{Puchkaryov1998} E. Puchkaryov and K. Maki, Eur. Phys. J. B. \textbf{4}, 191 (1998).
\bibitem{Ohashi2000} Y. Ohashi, J. Phys. Soc. Jpn. \textbf{69}, 2977 (2000).
\bibitem{Yoshioka2012} Y. Yoshioka and K. Miyake. J. Phys. Soc. Jpn. \textbf{81}, 093702 (2012).
\bibitem{note1} Effects of odd-frequency pairs formed near a vortex core on $T_1^{-1}$ have recently been discussed in Refs. \citen{Tanaka2016} and \citen{Tanaka2018}. However, $T_1^{-1}$ in the {\it bulk} odd-frequency superconducting state has not been examined yet.
\bibitem{Tanaka2016} K. K. Tanaka, M. Ichioka, and S. Onari, Phys. Rev. B \text{93}, 094507 (2016).
\bibitem{Tanaka2018} K. K. Tanaka, M. Ichioka, and S. Onari, Phys. Rev. B \text{97}, 134507 (2018).
\bibitem{Heid1995} R. Heid, Z. Phys. B \textbf{99}, 15 (1995).
\bibitem{Solenov2009} D. Solenov, I. Martin, and D. Mozyrsky, Phys. Rev. B \textbf{79}, 132502 (2009).
\bibitem{Kusunose2011B} H. Kusunose, Y. Fuseya, and K. Miyake, J. Phys. Soc. Jpn. \textbf{80}, 054702 (2011).
\bibitem{Fominov2015} Y. V. Fominov, Y. Tanaka, Y. Asano, and M. Eschrig, Phys. Rev. B \textbf{91}, 144514 (2015).
\bibitem{note} Reference~\citen{Fominov2015} points out that the path-integral approach still gives an unphysical result for tunneling current across a junction; however, we only deal with the bulk properties of the odd-frequency superconducting state, so that we do not meet this problem in this paper.
\bibitem{Stratonovich1958} R. L. Stratonovich, Soviet Physics Doklady \textbf{2}, 461 (1958).
\bibitem{Hubbard1959} J. Hubbard, Phys. Rev. Lett. \textbf{3}, 77 (1959).
\bibitem{Tempere2012} J. Tempere and J. P. Devreese, in {\it Superconductors materials, properties and applications}, ed. by A. Gabovich (IntechOpen, Rijeka, 2012) Chap. 16, p. 383.
\bibitem{Randeria2008} R. B. Diener, R. Sensarma, and M. Randeria, Phys. Rev. A \textbf{77}, 023626 (2008).
\bibitem{Stoof2009} H. T. C. Stoof, D. B. M. Dickerscheid, and K. Gubbels, Ultracold quantum fields (Springer, The Netherlands, 2009).
\bibitem{Iskin2005} M. Iskin and C. A. R. S\'a de Melo, Phys. Rev. B \textbf{72}, 224513 (2005).
\bibitem{Iskin2006A} M. Iskin and C. A. R. S\'a de Melo, Phys. Rev. Lett. \textbf{96}, 040402 (2006).
\bibitem{Iskin2006B} M. Iskin and C. A. R. S\'a de Melo, Phys. Rev. A \textbf{74}, 013608 (2006).
\bibitem{Iskin2006C} M. Iskin and C. A. R. S\'a de Melo, Phys. Rev. Lett. \textbf{96}, 040402 (2006).
\bibitem{Cao2013} G. Cao, L. He, and P. Zhuang, Phys. Rev. A \textbf{87}, 013613 (2013).
\bibitem{Tempere2008} J. Tempere, S. N. Klimin, J. T. Devreese, and V. V. Moshchalkov, Phys. Rev. B \textbf{77}, 134502 (2008).
\bibitem{note4} The superconducting phase transition temperature, of course, depends on the pairing type, so that one should write $T_{\rm c}$ in the $l$-type pairing as $T_{{\rm c},l}$. However, for simplicity, we always use the notation $T_{\rm c}$ irrespective of the pairing type, unless any confusion may occur.
\bibitem{Nagai2008} Y. Nagai, N. Hayashi, N. Nakai, H. Nakamura, M. Okumura and M. Machida, New J. Phys. \textbf{10}, 103026 (2008).
\bibitem{notez} We take $\delta/T_{\rm c}=10^{-2}$ in computations, in order to make the coherence peak just below $T_{\rm c}$ converge. Because of this prescription, the logarithmic singularity of $\rho_{\Delta,{\rm odd},s}(\omega)$ at $\omega=0$ also converges in Fig. \ref{fig4}(b).
\bibitem{Randeria1995} See, for example, M. Randeria, in {\it Bose-Einstein Condensation}, ed. by A. Griffin, D. W. Snoke, and S. Stringari (Cambridge University Press, New York, 1995), p. 355.
\bibitem{Altrand} A. Altland and B. D. Simons, \textit{Condensed Matter Field Theory}, (Cambridge University Press, New York, 2010), Chap. 7.
\end{thebibliography}
\end{document}